\documentclass[preprints,article,accept,pdftex,moreauthors]{Definitions/mdpi} 
\firstpage{1} 
\pubvolume{1}
\issuenum{1}
\articlenumber{0}
\pubyear{2025}
\copyrightyear{2025}
\datereceived{ } 
\daterevised{ } 
\dateaccepted{ } 
\datepublished{ } 
\hreflink{https://doi.org/} 

\Title{Human-Machine Collaboration-Guided Space Design: Combination of Machine Learning Models and Humanistic Design Concepts}

\TitleCitation{Title}

\Author{Yuxuan Yang $^{1,}$*}

\AuthorNames{Yuxuan Yang}

\isAPAStyle{%
       \AuthorCitation{Lastname, F., Lastname, F., \& Lastname, F.}
         }{%
        \isChicagoStyle{%
        \AuthorCitation{Lastname, Firstname, Firstname Lastname, and Firstname Lastname.}
        }{
        \AuthorCitation{Lastname, F.; Lastname, F.; Lastname, F.}
        }
}

\address{%
$^{1}$ \quad Nanjing Forestry University}

\corres{Correspondence: qingzhuo13@126.com}

\abstract{The integration of machine learning (ML) into spatial design holds immense potential for optimizing space utilization, enhancing functionality, and streamlining design processes. ML can automate tasks, predict performance outcomes, and tailor spaces to user preferences. However, the emotional, cultural, and aesthetic dimensions of design remain crucial for creating spaces that truly resonate with users—elements that ML alone cannot address. The key challenge lies in harmonizing data-driven efficiency with the nuanced, subjective aspects of design.
This paper proposes a human-machine collaboration framework to bridge this gap. An effective framework should recognize that while ML enhances design efficiency through automation and prediction, it must be paired with human creativity to ensure spaces are emotionally engaging and culturally relevant. Human designers contribute intuition, empathy, and cultural insight, guiding ML-generated solutions to align with users' emotional and cultural needs. Additionally, we explore how various ML models can be integrated with human-centered design principles. These models can automate design generation and optimization, while human designers refine the outputs to ensure emotional resonance and aesthetic appeal.
Through case studies in office and residential design, we illustrate how this framework fosters both creativity and cultural relevance. By merging ML with human creativity, spatial design can achieve a balance of efficiency and emotional impact, resulting in environments that are both functional and deeply human.
}


\begin{document}


\vspace{6pt}

\section{Introduction}

Space design \cite{broadbent2003emerging,carmona2019principles,kent2007creative} has traditionally been a human-centered process, driven by the creativity and intuition of architects and designers. These professionals rely on personal experiences, cultural insights, and emotional sensitivity to create spaces that are functional, aesthetically pleasing, and emotionally resonant. The success of space design lies in balancing form and function, ensuring that environments not only serve practical needs but also foster well-being and a sense of belonging \cite{maclean2020questions,yang2024diffdesign}. Whether designing homes, offices, or public spaces, the goal is to create environments that engage users on both functional and emotional levels.

Historically, design decisions have been rooted in human intuition and cultural understanding. Designers consider how colors, materials, lighting, and spatial organization influence emotions, behavior, and social interactions \cite{ching2018interior,kim2020stochastic}. For example, residential spaces aim to evoke warmth and security, while commercial spaces prioritize productivity and collaboration. Public spaces, such as museums or parks, often reflect cultural heritage and community values. In each case, the focus is on creating spaces that support both practical and emotional needs.

However, modern design challenges have grown more complex. Spaces are now multifunctional, adaptable, and influenced by urbanization, sustainability concerns, and technological advancements \cite{schafer2012machine,wang2023amsa}. Residential spaces often double as workspaces, while offices evolve into collaborative hubs. This complexity demands greater efficiency, sustainability, and adaptability, prompting designers to turn to technology, particularly machine learning (ML) \cite{irbite2021artificial,wang2024comprehensive}.
ML offers powerful tools to enhance the design process. It can analyze vast datasets to optimize layouts, predict energy consumption, and assess environmental performance \cite{ozisikyilmaz2008efficient,almaz2024future}. For instance, ML algorithms can generate energy-efficient designs or forecast how design elements affect user behavior \cite{racec2016computational,yang2024diffdesign}. Additionally, ML automates repetitive tasks, such as generating design alternatives or selecting materials, freeing designers to focus on creative refinement \cite{wang2023awesome}.

Despite these advantages, ML has limitations. While it excels at optimizing quantifiable metrics like energy efficiency and cost, it struggles to capture the emotional, cultural, and aesthetic nuances of design \cite{wu2022interior}. A purely data-driven approach may result in spaces that are functionally optimal but emotionally sterile or culturally disconnected. For example, an ML-generated design might maximize natural light but fail to consider how color schemes or spatial arrangements impact users' emotional well-being \cite{wang2024meta}. Cultural significance, such as the meaning of certain colors or materials, may also be overlooked.

To address these challenges, this paper proposes a human-machine collaboration framework \cite{tanasra2023automation,wu2022interior}. This approach leverages ML for data-driven optimization while ensuring that human creativity, cultural understanding, and emotional sensitivity remain central to the design process. ML can inform decisions on layout generation, material selection, and performance prediction, while human designers refine these outputs to ensure cultural relevance and emotional resonance.

Practical applications of this framework are diverse. In residential design, ML can optimize space efficiency and energy use, while human designers tailor layouts to reflect personal and cultural preferences \cite{yigit2021machine}. In office design, ML can analyze user behavior to enhance productivity, while human input ensures the space fosters collaboration and well-being \cite{wu2022intelligent}. Public spaces can benefit from ML's ability to optimize functionality while maintaining cultural and emotional significance \cite{zhuge2004resource}.
In conclusion, the future of space design lies in harmonizing technology and human creativity. By combining ML's computational power with human intuition and cultural insight, designers can create spaces that are both efficient and meaningful \cite{li2022multi}. This collaborative approach ensures that environments not only meet functional requirements but also resonate emotionally and culturally with their users, ultimately enhancing the human experience.

\section{Background and Motivation}

\subsection{Machine Learning in Space Design}

Machine learning (ML) has become an increasingly influential tool in spatial design, with the most promising applications including generative design, performance prediction, and personalized design, revolutionizing the way architects, interior designers, and urban planners create space.\cite{moosavi2020role}.  With the growing complexity of modern spaces, there is an increasing need for more efficient, adaptable, and data-driven design strategies. ML offers powerful algorithms that can optimize multiple aspects of space design, such as layout generation, material selection, energy efficiency, and user experience\cite{tien2022machine,kamalzadeh2022potential}. By leveraging large datasets and advanced modeling techniques, machine learning has the potential to automate and enhance design decisions, making it easier to achieve optimal functional and aesthetic outcomes.

\begin{figure*}[t]
    \centering
    \includegraphics[width=\textwidth]{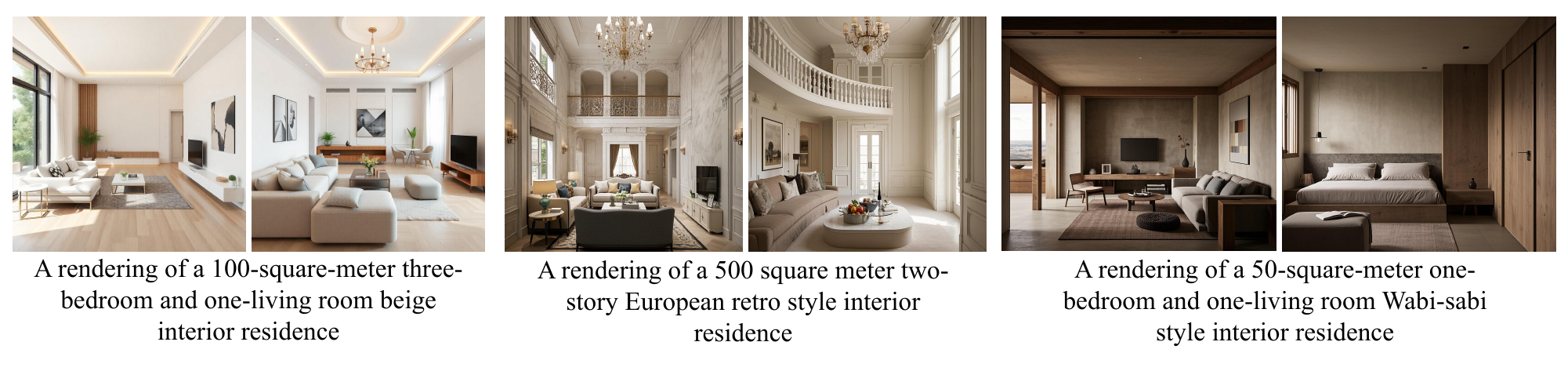}
    \caption{Generative design creates different renderings of interior spaces based on home size, number of rooms, or specific preferences for room configurations.}
    \label{fig:case_1}
\end{figure*}

Generative design refers to algorithms like generative adversarial networks (GANs) and evolutionary algorithms, which can create a wide variety of design alternatives based on user-defined parameters and constraints\cite{yuan2022deep}. These algorithms explore the solution space by considering factors such as spatial constraints, functional requirements, aesthetic preferences, and material properties. By generating multiple iterations of a design, generative design allows designers to consider various options quickly, making it easier to identify the most suitable layout and design choices for a specific project. For example, in residential design, generative design can create a range of floor plans based on the size of the family, the number of rooms, or specific preferences for room configuration (Figure \ref{fig:case_1}).

    \begin{figure*}[t]
    \centering
    \includegraphics[width=\textwidth]{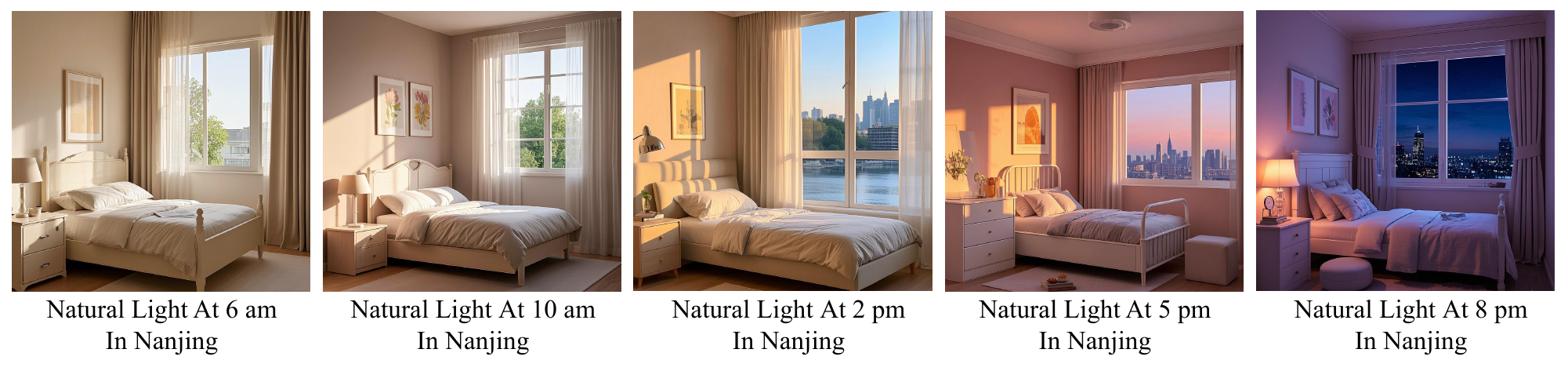}
    \caption{ML algorithms predict how natural light enters spaces throughout the day in Nanjing.}
    \label{fig:case_2}
\end{figure*}

\begin{figure*}[t]
    \centering
    \includegraphics[width=\textwidth]{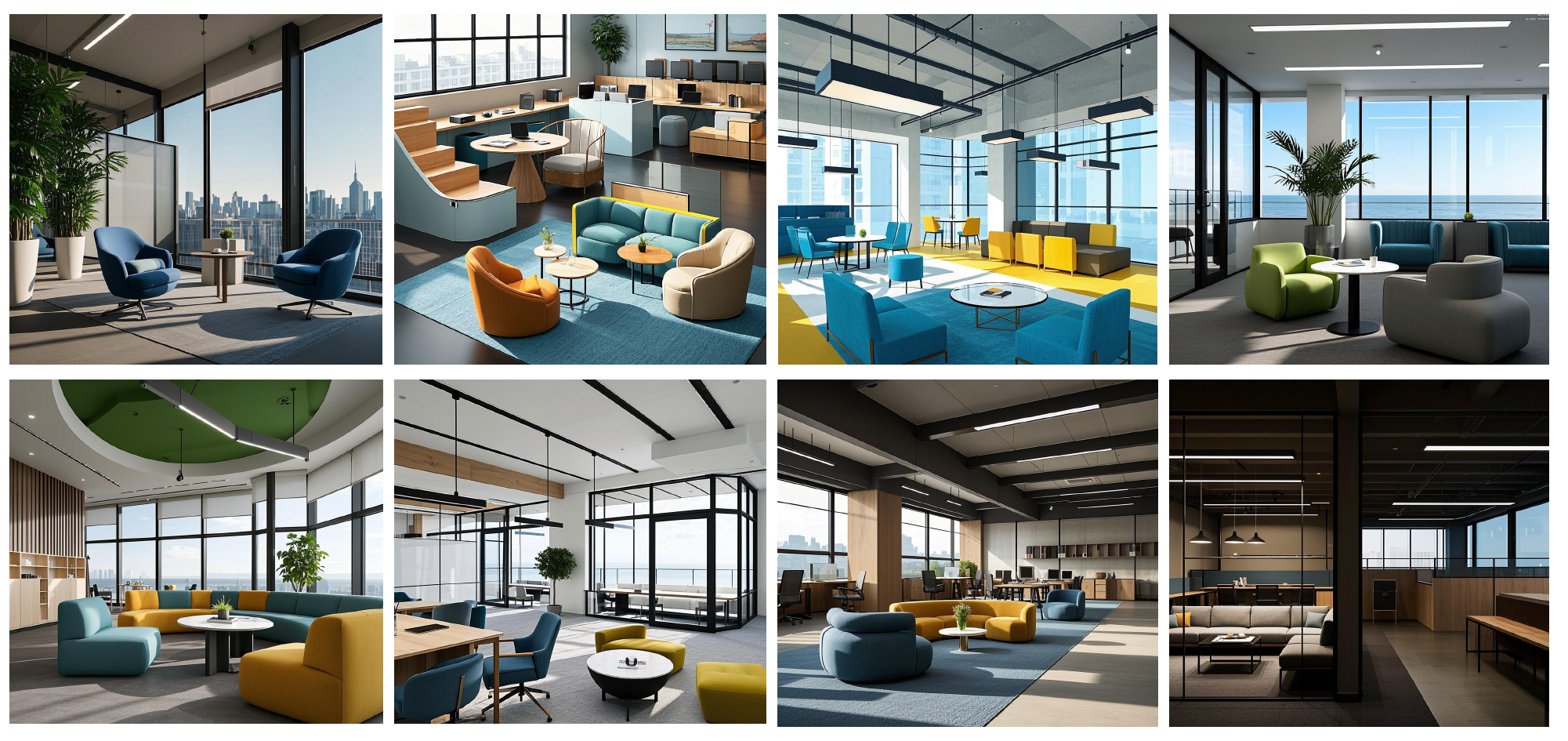}
    \caption{ML optimizes the renderings of the casual office space in the office space and gives it different lighting effects.}
    \label{fig:case_3}
\end{figure*}

Machine learning models can be employed to predict the environmental and operational impact of various design choices\cite{lirecent,wang2024towards}. These models simulate the effects of different materials, layouts, and environmental factors such as lighting, temperature, and airflow. For instance, ML algorithms can predict how natural light will enter a space throughout the day(Figure \ref{fig:case_2}), assess how various layouts influence airflow patterns, or calculate energy consumption based on the materials used. These predictions help designers make informed decisions that improve the overall sustainability, comfort, and energy efficiency of a building. Predictive analytics can also assist in assessing the long-term performance of a design, helping to anticipate potential issues such as heat loss or poor acoustics. Meanwhile, Machine learning can also enable designers to tailor spaces to individual needs and preferences. By analyzing user data, such as behavioral patterns, feedback, and demographic information, ML models can suggest design elements that align with users' personal preferences or cultural expectations\cite{wang2024machine}. In residential design, this could mean adjusting room layouts based on the size of the family or customizing furniture and color schemes to reflect the homeowners’ tastes. In commercial or office spaces, ML can optimize layouts to promote employee well-being and productivity, such as adjusting lighting to reduce stress or designing collaborative spaces to encourage creativity(Figure \ref{fig:case_3}). Personalization allows designers to create spaces that are not only functional but also resonate with the users' unique identities and needs.

Despite these advantages, most machine learning models in design are primarily focused on functional and performance metrics. These models excel in optimizing for efficiency, cost, and operational functionality, providing valuable insights for creating high-performance spaces. However, while these aspects are essential, they often overlook the emotional and cultural dimensions that are critical to creating truly engaging and meaningful spaces. A design that is optimized for energy efficiency or cost-effectiveness may still lack the emotional connection that users need to feel comfortable and at home in the space.

\subsection{Humanistic Design Principles}

Humanistic design principles emphasize the importance of empathy, cultural sensitivity, and an in-depth understanding of human behavior and needs. Aesthetic harmony, cultural relevance, and user comfort are key elements of humanistic design. Rather than merely focusing on functional outcomes, humanistic design seeks to create spaces that promote psychological and emotional well-being, foster social interaction, and reflect the cultural and social contexts of the people who use them\cite{yang2011study}. This approach is based on the belief that design should not just serve practical purposes but should also enhance users' lives by providing environments that resonate with their emotions, values, and cultural identities.

\begin{figure*}[t]
    \centering
    \includegraphics[width=\textwidth]{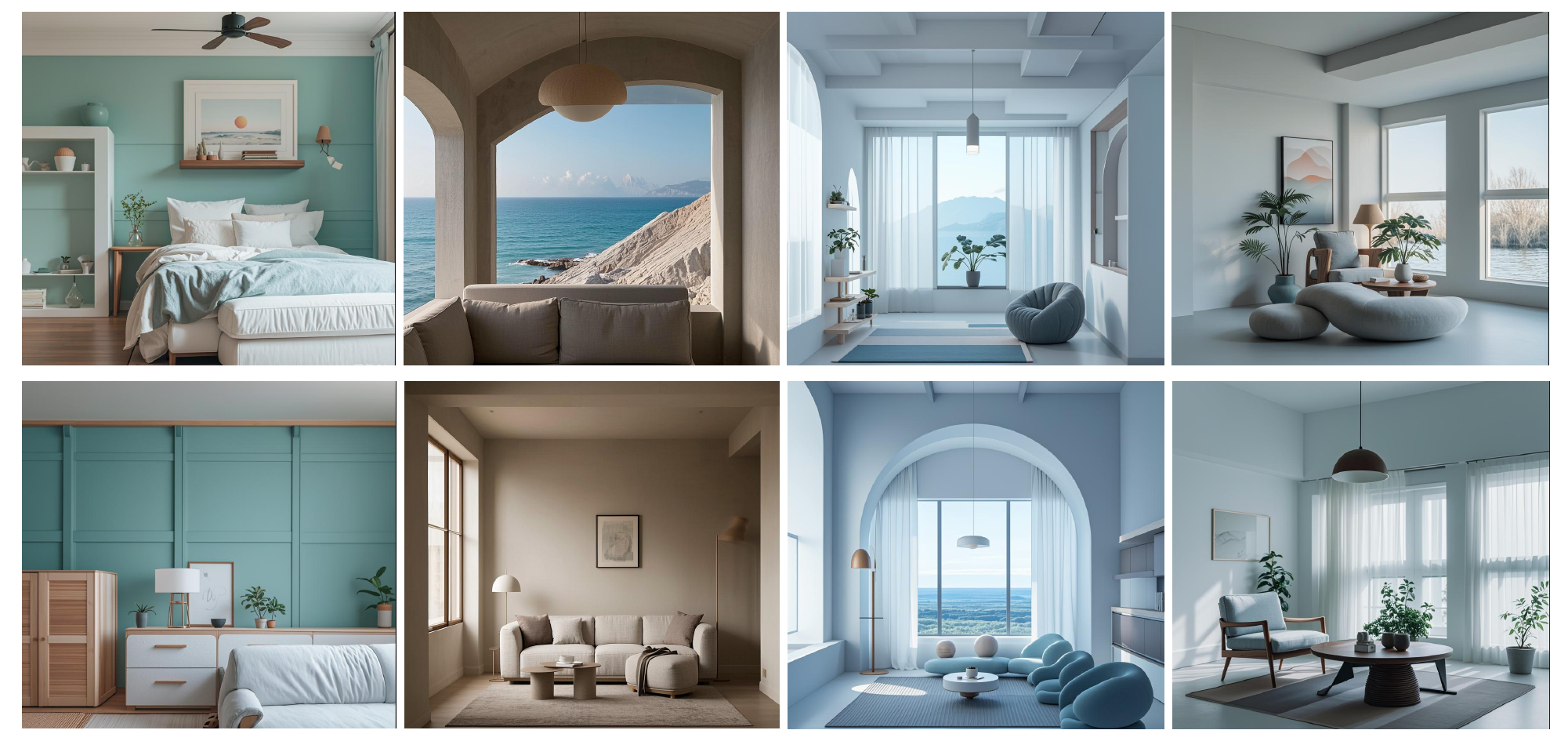}
    \caption{ML models can help create spaces with calming tones, balanced proportions, and natural textures that promote emotional well-being and reduce stress.}
    \label{fig:case_4}
\end{figure*}

Aesthetic harmony in design refers to creating environments that evoke positive emotional responses through the careful selection of colors, forms, textures, and layouts. By paying attention to the visual and sensory qualities of a space, designers can create an atmosphere that fosters peace, relaxation, and inspiration\cite{moon1944aesthetic,schloss2011aesthetic}. For example, calming color palettes, well-balanced proportions, and natural textures can help create spaces that promote emotional well-being and reduce stress (Figure \ref{fig:case_4}). This aspect of design is particularly important in spaces like bedrooms or wellness centers, where comfort and tranquility are paramount.

Cultural relevance is another cornerstone of humanistic design. Spaces should be designed with an understanding of the cultural backgrounds and values of the people who will occupy them\cite{pelletier2012space}. By incorporating cultural symbols, traditions, and preferences, designers ensure that the space resonates with its users on a deeper, more personal level. This might involve incorporating local art, honoring historical or cultural references, or using materials that are meaningful to a specific community. In a multicultural society, creating culturally relevant spaces helps promote inclusivity, understanding, and a sense of belonging.

\begin{figure*}[t]
    \centering
    \includegraphics[width=\textwidth]{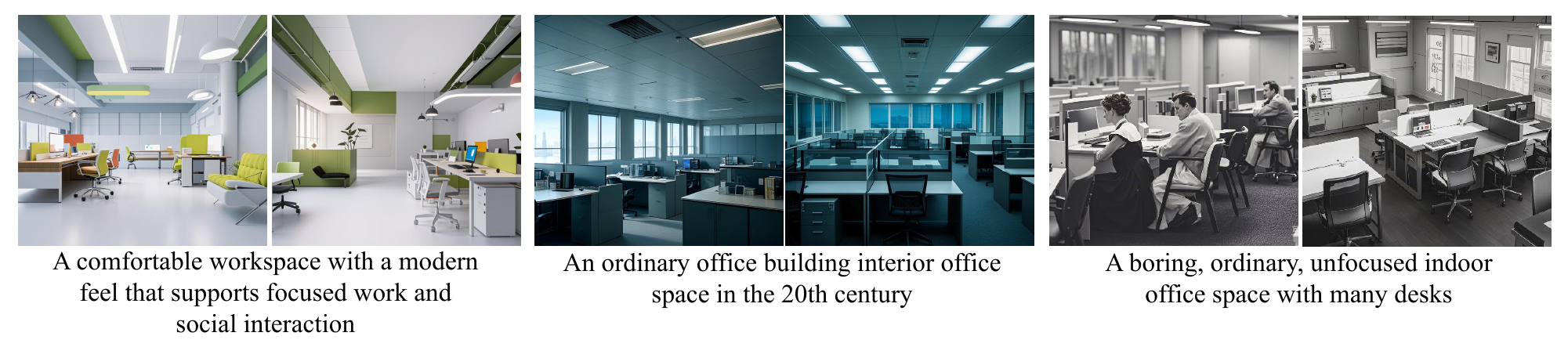}
    \caption{ML can create comfortable spaces that are more supportive of focused work and social interaction, thereby increasing productivity and creativity.}
    \label{fig:case_5}
\end{figure*}

Humanistic design also prioritizes user comfort, which goes beyond just functional needs. This involves optimizing the physical environment to enhance both psychological and physiological comfort. For example, designers consider factors such as lighting quality, acoustics, ergonomics, and thermal comfort. Additionally, user comfort involves creating spaces that foster emotional well-being by considering how design elements affect mood and behavior\cite{zhou2024health}. In workplaces, for instance, a comfortable space can increase productivity and creativity by offering a balanced environment that supports both focused work and social interaction(Figure \ref{fig:case_5}).

Humanistic design, however, is not without its challenges. It is inherently subjective, and as such, there is no universal approach to creating emotionally resonant or culturally relevant spaces. Emotional responses to design elements can vary significantly between individuals based on their personal preferences, cultural backgrounds, and life experiences\cite{ho2012emotion}. Furthermore, the intangible and subjective nature of human well-being makes it difficult to quantify or standardize in the same way that functional metrics, such as energy efficiency, can be measured. This subjectivity poses a challenge for integrating humanistic principles into algorithmic design processes, where design decisions are often based on objective, data-driven criteria. As a result, finding a way to balance the data-driven nature of machine learning with the emotional and cultural sensitivity required in humanistic design is a critical challenge that must be addressed for successful space design.
By bridging the gap between the objective, performance-based strengths of machine learning and the subjective, emotion-driven elements of humanistic design, it is possible to create spaces that are both highly functional and deeply meaningful to the people who use them. This intersection is where the true potential of human-machine collaboration in space design lies.

\section{Methodology: Human-Machine Collaboration Framework}

\subsection{Integrating Machine Learning Models into Design Processes}

To successfully integrate machine learning into space design while maintaining humanistic principles, we propose a collaborative framework that balances the efficiency of ML with the emotional and cultural needs of the users. 
The primary function of the machine learning component within the framework is to generate design alternatives that optimize space utilization, functionality, and performance based on predefined criteria. These criteria can include a wide range of factors such as spatial efficiency, lighting optimization, energy consumption, ventilation, and acoustics\cite{hussein2025metaheuristic}. The system leverages large datasets from past designs, user preferences, and architectural performance metrics to propose various design solutions that meet functional goals.
Machine learning algorithms such as supervised learning, reinforcement learning, and generative models (e.g., GANs or VAEs) are utilized to explore a large design space and generate diverse layout configurations. The ML model is trained on historical design data, user interaction patterns, and environmental factors to generate layouts that are not only efficient but also sustainable\cite{shoushtari2024facilities,hussein2024machine}.
For example, the system might suggest a design where the kitchen is located near natural light sources, or a room configuration that optimizes the flow of people based on typical usage patterns\cite{uddin2025predicting}. Once the system proposes a range of alternative designs, human designers can review and select the most promising options, which can then be further refined.

This ML-driven approach significantly enhances the efficiency of the design process by automating the generation of space layouts and considering a wider set of performance metrics, enabling designers to focus on creative customization rather than manually exploring each design possibility.
The core components of this framework are:

\begin{itemize}
    \item \textbf{Machine Learning-Driven Design Suggestions}: The system generates multiple design alternatives based on specific functional and performance criteria. These designs are optimized based on parameters such as space usage, lighting, ventilation, and energy consumption. Designers can select from these options or modify them based on their creative vision.
    \item \textbf{Human Feedback Loop}: After generating initial designs, designers can provide feedback on the machine-generated outputs. This feedback loop allows designers to refine the design based on human insights and preferences, ensuring that the final outcome reflects humanistic values such as cultural relevance, emotional resonance, and user comfort.
    \item \textbf{Interactive Design Environment}: The framework includes an interactive platform where designers can adjust the machine-generated designs in real time, adding personal touches or modifying features to align with aesthetic preferences, ergonomic needs, or cultural expectations.
    \item \textbf{Ethnographic and Cultural Analytics}: A specialized module analyzes data on cultural symbols, historical context, and social behaviors to offer suggestions that ensure the designs are culturally sensitive and relevant. By drawing on large datasets, this module can suggest design elements that are in line with local traditions and values.
\end{itemize}
Each of these components works together to create a collaborative design process where machine learning models enhance the efficiency and performance of the design, while human designers provide the necessary intuition, empathy, and cultural sensitivity that make the design truly human-centered. Through iterative feedback, real-time interaction, and cultural analysis, the framework ensures that the design outcomes are both innovative and aligned with humanistic values, ultimately resulting in spaces that are functional, emotionally engaging, and culturally meaningful.

\subsubsection{Human Feedback Loop}

While machine learning algorithms excel at generating design alternatives based on data-driven insights, they often fall short in capturing the subtle human factors that can make a design truly successful\cite{camburn2020machine,kaluarachchi2021review}. These factors, such as emotional resonance, comfort, and cultural relevance, are critical to creating spaces that truly meet user needs. To address this limitation, the framework proposed in this paper embeds a feedback loop that enables human designers to directly participate in the machine-generated design process, incorporating their unique insights and expertise to ensure that the design is not only optimal in terms of function, but also emotionally and culturally in line with user expectations.

Once the machine learning system generates initial design alternatives, human designers begin the evaluation process. This evaluation goes beyond simply assessing functionality; designers also consider emotional aspects and user comfort\cite{sun2020developing}.For example, designers might adjust the layout of a space to create a harmonious atmosphere, optimize the orientation of a space to maximize natural light, and modify the layout of furniture to promote social interaction among users\cite{wu2020product}. These adjustments reflect the designer’s understanding of the intangible elements of space that are critical to creating an environment that users feel close to.

This feedback is not a one-time interaction; rather, it forms an iterative process where human insights are continuously fed back into the system. As designers provide input, the machine learning model learns from these interactions, refining its design suggestions based on user preferences\cite{gupta2021research}. For example, when a designer suggests improving traffic flow or adjusting lighting, the system adjusts and generates a revised design in real time. Over time, the system gradually learns the nuances of human preferences, making the design process more personalized. The iterative nature of this feedback loop ensures that each new design iteration builds on the previous one and is continuously improved based on human feedback. This not only improves the technicality of the design, but also strengthens its alignment with human values such as comfort, emotional appeal, and cultural relevance\cite{li2021survey}. As the system continues to learn, it incorporates more complex emotional and social cues, thereby creating spaces that resonate more deeply with users.

Moreover, the ability to integrate human feedback in real-time allows the design process to be highly dynamic and adaptive\cite{mccormack2020design}.Designers are not limited by static outputs and can actively shape the design based on their insights and experiences to ensure it meets the emotional and psychological needs of users, promoting a deeper connection with the space. For example, in home design, designers may request the addition of elements such as soft textures or private seating to create a warm and comfortable space, and the system will adjust the plan based on feedback to provide design solutions that reflect a human-centric approach.

\begin{figure*}[t]
    \centering
    \includegraphics[width=\textwidth]{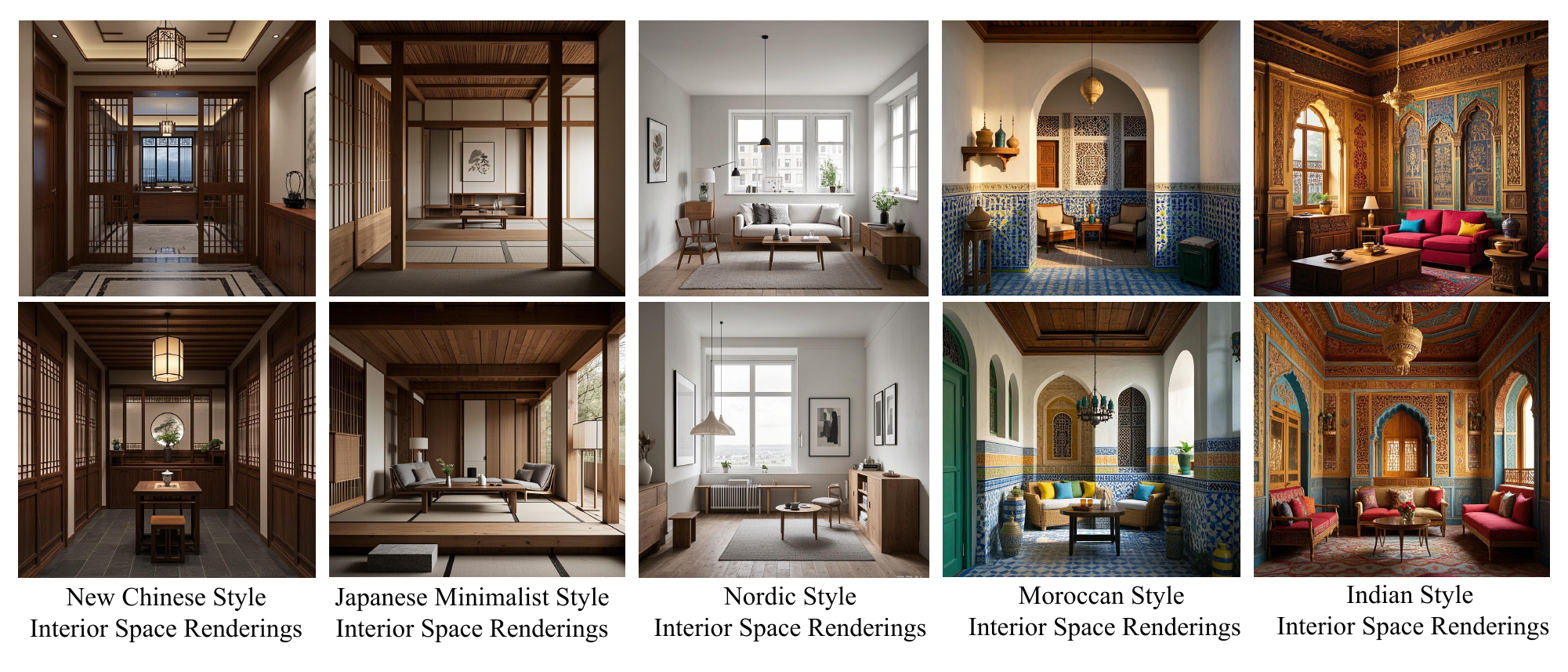}
    \caption{ML can generate designs that are both functionally sound and culturally sensitive, ensuring that the final design is meaningful and appropriate for the context in which it will be used.}
    \label{fig:case_6}
\end{figure*}

In addition to personal preferences, designers also incorporate their cultural and contextual understanding into the design process. Designers working on projects in a specific cultural context may use special materials or patterns to enhance the emotional and cultural significance of the design\cite{julier2013culture}. Machine learning systems learn from these inputs and can then generate designs that are both functional and culturally sensitive, ensuring they are appropriate for the context in which they are used(Figure \ref{fig:case_6}).

By integrating human feedback, machine learning models can collaborate with designers to reflect the complexity of human needs. This approach balances technical efficiency with human emotional, social, and cultural insights to create efficient and human-centered spaces\cite{winter2020flatpack,buschek2020paper2wire}. Ultimately, the feedback loop makes the design process more dynamic, adaptive, and responsive, with each iteration bringing the design closer to a combination of function and emotion. As the system continues to learn, it will become better at predicting and anticipating future needs, and its design recommendations will more accurately predict user needs, achieving a harmonious fusion of technology and creativity.

\subsubsection{Interactive Design Environment}

The interactive design environment plays a central role in bridging the gap between machine-generated designs and human creativity in the space design process\cite{sindiramutty2025generative}. It provides designers with the tools and features needed to manipulate and refine machine-generated layouts in real time. This dynamic interaction allows for the seamless integration of human insights, ensuring that design decisions are both data-informed and emotionally resonant\cite{costa2024artificial}.

The core of the platform lies in its intuitive operation tools, which allow designers to adjust the positions of key elements such as walls and furniture in real time through functions such as drag and drop. This simple operation method not only simplifies the process of trying various configurations and speeds up the layout adjustment process, but also inspires designers to explore the potential of diversified design solutions, making the design process more flexible and efficient.

The interactive platform also offers 3D visualization and interactive floor plans to provide designers with an immersive design experience. These tools can help designers more intuitively understand how their choices will manifest in the real world and better understand scale, proportion, and spatial relationships, which are often difficult to evaluate through 2D-floor plans alone. By understanding how elements fit together, designers can make more informed decisions about how to optimize space, lighting, and circulation. For example, a designer may adjust the layout of a room and immediately see how the new configuration affects the atmosphere or function of the space, making it easier to balance aesthetic preferences and practical limitations, and deepening the understanding of spatial proportions and relationships. These personalized touches elevate the design from a purely functional output to a solution tailored to the client's unique needs.

\begin{figure*}[t]
    \centering
    \includegraphics[width=\textwidth]{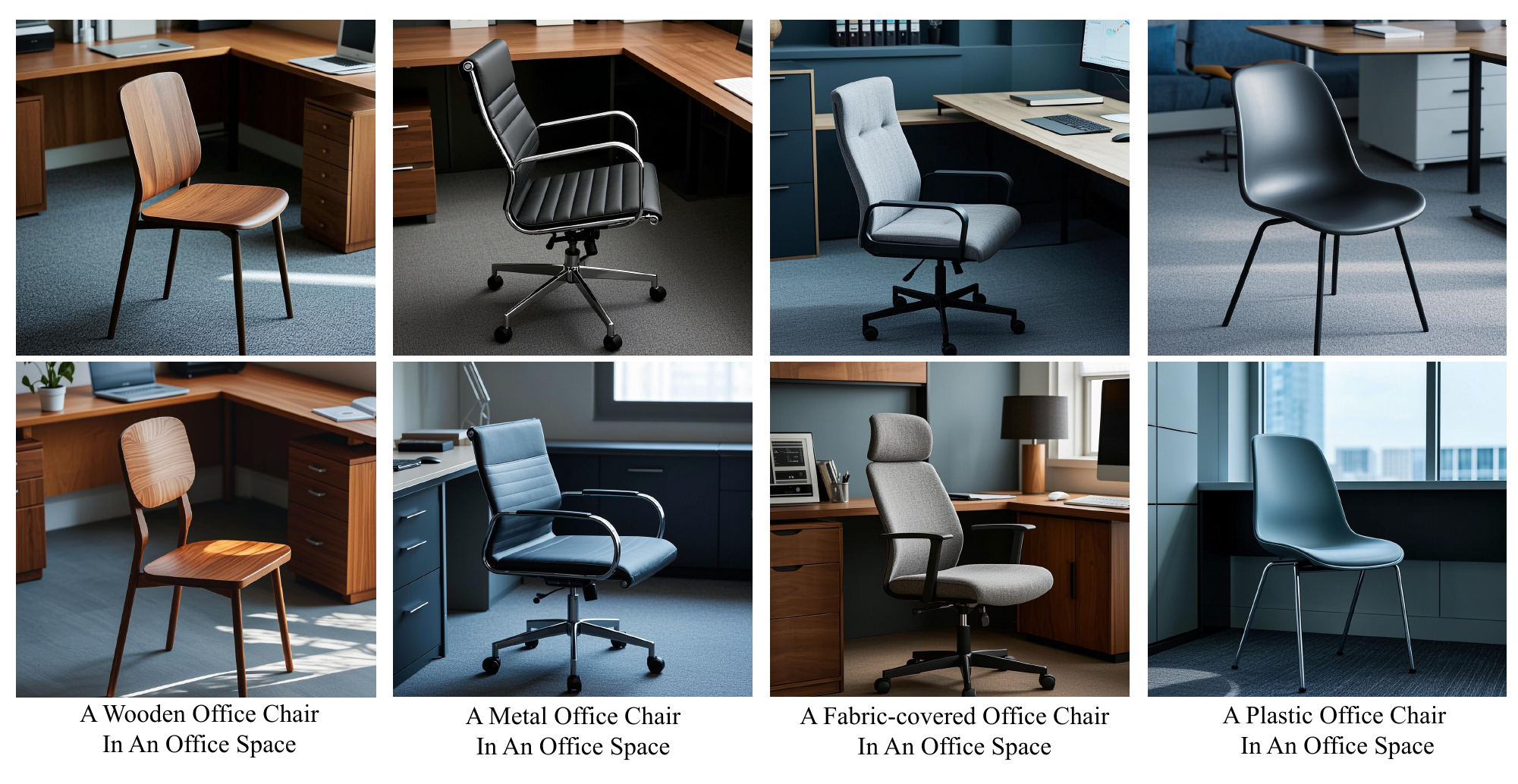}
    \caption{ML can try generating different materials (such as wood, metal, or fabric) to see how each affects the ambiance of a room.}
    \label{fig:case_7}
\end{figure*}

The integration of environmental simulation capabilities further enhances the depth of the design process. Designers can simulate spatial performance under different conditions, enabling the exploration of multiple “what-if” scenarios. By testing different combinations of materials, furniture, and spatial arrangements, their impact on spatial functionality and emotional experience can be evaluated. For example, designers can try different materials (such as wood, metal, or fabric) to understand how each material affects the atmosphere of a room to determine which layout can maximize comfort and promote social interaction\cite{baduge2022artificial}(Figure \ref{fig:case_7}). This real-time and flexible performance evaluation can help designers find the best balance between aesthetics and functionality, enabling designers to continuously iterate and improve designs until the optimal design is reached, without being constrained by initial assumptions or limitations\cite{peters2020product}.

Interactive design environments also promote a more dynamic and collaborative design process\cite{olsson2020technologies}. The real-time interaction between the designer and the machine learning system forms a continuous feedback loop, where each adjustment made by the designer allows the system to learn and optimize future suggestions, making the system more intelligent and adaptive to the designer’s needs. If a designer repeatedly adjusts a layout to emphasize open space, the system may begin to recommend designs that prioritize spatial openness in future iterations\cite{jacobs2022furthering}. At the same time, it allows for rapid communication, real-time feedback, and decision-making among all stakeholders, including clients, architects, and interior designers, ensuring that the final design meets the client’s expectations. The collaboration between human experts and machine learning systems creates a truly collaborative design approach, where technology supports human creativity while also benefiting from human intuition, cultural awareness, and aesthetic judgment.

By supporting such an open, collaborative, and flexible design process, the interactive design environment plays a pivotal role in ensuring that the final designs are both functional and human-centered\cite{xiang2020sapien}. It provides designers with the ability to manipulate, visualize, and optimize machine-generated designs in real-time, while machine learning models provide data-driven insights to improve design efficiency and performance, ensuring that the final design is both practical, personalized, and culturally appropriate. As the design process becomes increasingly iterative and collaborative, the role of interactive platforms will continue to grow, driving design solutions towards more adaptive, responsive, and user-centric directions\cite{boukhelifa2020challenges}.

\subsubsection{Ethnographic and Cultural Analytics}

The core highlight of the proposed framework is that it integrates ethnography and cultural analysis modules, which significantly promotes the concept of human-centered design. Through big data analysis and deep cultural insights, this module ensures that the design is not only functionally efficient, but also resonates with users at the cultural and emotional levels. With rich ethnographic data (covering cultural symbols, historical background, social behavior, and user preferences), the system can generate design suggestions that are highly consistent with specific cultural or regional contexts\cite{ploder2021practices,burns2020s}.

This cultural sensitivity is key to successful design\cite{spence2020senses}. For example, the system can recommend design elements that have an emotional connection or a regional identity based on cultural trends or historical context. For example, the color red, which symbolizes prosperity in a particular culture, may be suggested for use in certain design details. In addition, the system may recommend the use of local materials to promote sustainability and enhance connection with the environment.

The ethnography module can also take into account cultural preferences for spatial organization. By analyzing big data related to social behavior and cultural customs, the system can ensure that the design not only conforms to the values of individual users, but also takes into account broader social norms. For example, in traditional Eastern families, family dynamics and social behavior have a significant impact on the spatial design of the home\cite{pink2020making}. The system may recommend placing public areas in the center and private spaces on the periphery to reflect family structure and social customs. In cultures that value public life and social gatherings, the system will emphasize the design of public areas to promote interaction. By targeting different needs for privacy and public space in different regions, the system will make corresponding layout suggestions to ensure that the space is both comfortable and in line with user habits.

Cultural analysis modules also play a vital role in addressing design challenges in multicultural or global contexts\cite{plocher2021cross}. By analyzing datasets spanning different cultures, social classes, and historical backgrounds, the system can provide design recommendations that resonate with the cultural values of multiple user groups, which is particularly important in environments such as international offices, multiethnic communities, or hotels that cater to global tourists. For example, in the design of a multicultural office space, the system may recommend the use of a mix of furniture styles or layouts to ensure that the space feels welcoming and inclusive to people from different cultural backgrounds in order to foster a sense of belonging among employees. In addition, the module can provide valuable insights into the inheritance and evolution of architectural styles in a particular region by analyzing historical design patterns such as vernacular architecture. For example, the modern design of a building in a historic district may incorporate elements such as local stone facades, traditional rooflines, or other architectural details that echo the past, creating an effect that seamlessly blends the old and the new.

Another compelling aspect of the Ethnography and Cultural Analysis module is its ability to assess the social impact of design decisions. The module can evaluate how different design elements affect the social behavior of occupants. For example, in public spaces such as community centers, the system may recommend designing flexible seating arrangements or public art areas to promote interaction and community engagement, while in private residential designs, it may prioritize quieter, more intimate spaces to support family connections and personal privacy.

By incorporating cultural context, historical significance, and social behavior into the design process, the ethnography and cultural analysis module not only enhances the functionality and aesthetics of the design, but also gives the space deeper cultural meaning and emotional value. This approach not only strengthens the user’s connection to the environment, but also fosters a sense of belonging and pride in the user. As design becomes increasingly interdisciplinary and global, this module will play an increasingly important role in creating spaces that respect human diversity\cite{al2021impact}.

\subsection{Enhancing Human-Centric Design Integration}

While machine learning models provide valuable insights, human designers still play an indispensable role in ensuring that designs are in line with humanistic principles. The human-centered design integration process follows these steps:

User-centered evaluation: Designers and customers can rate machine-generated layouts and provide detailed feedback, which enables the system to learn from human preferences and improve future suggestions. This ensures that the design process remains collaborative, avoids the algorithm-dominated singularity, and makes the design closer to the actual needs and emotional experience of users.

Iterative design improvement: Based on human input, machine learning models iteratively improve design suggestions. Ensuring that the design solution is both data-based and fully reflects humanistic care, this iterative process not only improves the efficiency of design, but also gives the space emotional appeal and cultural resonance.
Personalized algorithms: By analyzing user preferences, lifestyles, and habits, machine learning models can generate designs that reflect personal tastes and optimize space use. This personalization is not only reflected in functionality, but also through emotional design language, making the final space both practical and emotionally resonant for users.

Through the organic combination of user-centered evaluation, iterative design improvement, and personalized algorithms, the design process achieves a deep integration of technology and humanity. This collaborative model ensures that the design is not only efficient, but also emotionally engaging and culturally appropriate, creating truly human-centric spatial solutions.

\subsubsection{User-Centered Evaluation}

User-centered evaluation is a fundamental principle of human-centered design, which integrates user feedback and experience into the design process to ensure that solutions are not only functional and efficient, but also fit the user's emotional, cultural, and lifestyle preferences\cite{gottgens2021application}. In interior design, this approach transforms the traditional linear design workflow into a dynamic, interactive, and collaborative process. By leveraging human-computer interaction and machine learning models, designers and clients can conduct continuous evaluations, resulting in solutions that more accurately reflect human needs and preferences.

Taking Autodesk Revit as an example, the platform supports collaborative workflows in a shared digital environment, allowing users to provide real-time feedback while machine learning algorithms optimize layouts based on input\cite{stine2023design}. This dynamic interaction enables designers and clients to directly influence design decisions, such as automatically adjusting room layouts. However, Revit’s reliance on predefined templates limits creativity for complex or highly customized designs. Incorporating more advanced machine learning models can further enhance user-centered design capabilities and provide more personalized and adaptive solutions.

From a human-computer interaction perspective, intuitive and efficient interfaces are essential to facilitate user-centered evaluation. Platforms such as SketchUp provide a flexible 3D design environment and a user-friendly interface to help clients visualize spaces and provide feedback. However, SketchUp lacks advanced machine learning integration to dynamically adjust and generate new design iterations based on real-time user input. While designers and clients can collaborate to modify designs, the lack of real-time adaptive learning limits the platform's responsiveness.

Throughout the design process, machine learning models generate multiple layout options that designers and clients evaluate using an interactive interface. Platforms such as Revit and SketchUp support a variety of feedback mechanisms, including visual annotations, textual comments, and rating systems (e.g., a scale of 1-5), which enhance the model’s ability to interpret user preferences\cite{waas2022review}. For example, Revit allows users to annotate areas for improvement or suggest modifications to materials and layouts. By collecting and analyzing this feedback, machine learning models can optimize spatial configurations, improve functional flows, and better align designs with users’ aesthetic preferences.

Machine learning plays a key role in this evaluation process, particularly its ability to identify patterns in user feedback and improve the design accordingly. For example, if multiple users express dissatisfaction with lighting conditions, the model should detect this trend and adjust the lighting strategy in subsequent iterations. While Revit incorporates some of these adaptive features, its real-time learning capabilities are still limited. In contrast, platforms such as RoomSketcher provide convenient tools for creating and visualizing floor plans, but lack machine learning-driven feedback loops and are therefore less effective in dynamically optimizing designs.

From a design theory perspective, user-centered evaluation transforms the design process from relying solely on designer intuition to a more systematic, evidence-based approach. By combining subjective user judgment with scientific methods, this approach ensures that designs can be continuously optimized based on real user input, thereby enhancing their functional effectiveness and emotional resonance.

\subsubsection{Iterative Design Refinement}

Iterative design improvement is a dynamic process that combines continuous human feedback with machine learning models to produce adaptive, responsive, and personalized spaces\cite{xie2020learning}. At its core, it leverages a circular feedback loop to ensure that user preferences, cultural nuances, and dynamic needs actively influence the design at every stage. This approach balances data-driven optimization with human-centered insights to create spaces that are functional, emotionally engaging, and culturally relevant.

Machine learning models play a key role in this process, responding to user input by continuously generating, optimizing, and re-evaluating design suggestions. For example, in residential design, the model can generate an initial layout based on user-defined parameters (such as the number of rooms, openness, and functional zoning), and then iteratively optimize it based on user feedback on spatial flow, lighting, or furniture placement, allowing the model to integrate each round of input and produce increasingly sophisticated and personalized solutions. By integrating quantitative indicators (such as spatial efficiency and energy consumption) with qualitative preferences (such as emotional and aesthetic needs), the final design achieves the best balance between functionality and personalization\cite{cong2022machine}.

This iterative process is particularly effective in solving complex, multi-dimensional design challenges that are often difficult to address with traditional approaches. For example, when designing for multicultural communities, personalization is not limited to functional considerations such as accessibility and budget constraints, but also includes cultural expectations for privacy, social interaction, and spatial organization. Through iterative improvements, the system can This iterative approach is particularly effective in solving complex, multi-dimensional design challenges. For example, when designing for multicultural communities, the system can not only meet functional requirements such as accessibility and budget, but also prioritize culturally relevant preferences for privacy, social interaction, and spatial organization, thereby learning to prioritize design elements that are consistent with the user's cultural context, ensuring that the final space resonates with their values and life experiences.

Additionally, iterative design improvements enable the flexibility to adapt to dynamic changes in user needs. If a user initially prefers an open layout, but later realizes the need for private areas due to changing lifestyle needs, the system can quickly adjust the design. This flexibility enabled by machine learning surpasses traditional design methods in terms of responsiveness and efficiency, ensuring that spaces remain relevant and adaptable over time, and rapid adjustments can minimize resource expenditures while keeping the design aligned with the user's current needs.

Beyond functional optimization, iterative refinement fosters deeper emotional connections between users and their spaces. Design is not merely about spatial arrangement; it is about crafting environments that evoke a sense of belonging, comfort, and identity\cite{matkovic2023iterative}. For instance, a system can analyze user preferences for colors, textures, and lighting to create atmospheres that support desired emotional responses—whether warmth and hospitality or tranquility and focus. The iterative process allows the design to evolve in response to these emotional considerations while maintaining practical functionality.  

A prime example of this approach in practice is generative design platforms such as Autodesk Generative Design, which leverages artificial intelligence (AI) and machine learning to generate optimized solutions based on user-defined goals and constraints. These systems continuously refine designs by incorporating feedback, demonstrating the power of iterative design refinement in achieving solutions that are both computationally optimized and deeply attuned to human needs.

\subsubsection{Personalization Algorithms}

Personalization algorithms are at the core of modern design developments, and they are able to customize spaces based on users’ unique preferences, lifestyles, and habits. These algorithms use machine learning to analyze large amounts of user data (e.g., interactions with design elements, daily life, spatial behavior, and personal tastes) to generate designs that are both functionally optimized and emotionally engaging, thereby achieving user-centered design that allows spaces to both meet practical needs and reflect personal identity\cite{rafieian2023ai}.

At the heart of these algorithms is the ability to interpret and translate subjective user preferences into specific design features, significantly enhancing the emotional appeal of the design. Human emotional responses are closely tied to the surrounding environment, and personalized design can have a profound impact on feelings of well-being. For example, a personalized algorithm may identify that a user prefers open, airy spaces filled with natural light and minimalist furniture, or layouts that favor social gatherings. Based on these insights, the system recommends larger windows, flexible open spaces, and reconfigurable seating arrangements. By analyzing these preferences, the system can create an environment that meets emotional needs and generate designs that create a sense of comfort, harmony, and belonging, ensuring that the space is not only practical but also emotionally resonant.

Personalization algorithms also optimize space usage by observing how users interact with their environment. By analyzing patterns such as movement, frequency of space use, and preferences for spatial layout, the system can recommend layouts that maximize available space, eliminate inefficiencies, and improve overall convenience\cite{yoganarasimhan2020search}.For smaller apartments, the system might recommend multifunctional furniture to create a more flexible living space. As the system collects more data about user behavior over time, it will continually optimize the design to adapt to the user’s dynamic needs.

Incorporating personalized algorithms into the design process can foster a deeper level of collaboration between humans and machines. Traditional interior design relies on designers’ interpretation of user needs, but with personalized algorithms, users become active participants, shaping the design based on their personal tastes and requirements. This approach can foster a greater sense of ownership and satisfaction, as users see their preferences reflected in the final space, leading to a more meaningful and fulfilling design experience.

A key advantage of personalization algorithms is their ability to process large, complex data sets that are often beyond the reach of traditional approaches. By aggregating and analyzing data from multiple sources (e.g., user interactions, social media activity, and historical design experience), the system can generate highly customized designs that balance functionality and integrity\cite{schjott2023designing}. In commercial spaces, algorithms can adjust layouts based on employee preferences, team dynamics, or even emotional patterns detected by smart sensors to optimize productivity, collaboration, or customer experience.

However, personalization algorithms also face privacy and ethical challenges. The system relies on personal data to make design decisions, and strict measures must be taken to protect sensitive information and ensure user trust. In addition, while the algorithm is designed to capture personal preferences, it must recognize that tastes and habits will develop over time. Therefore, it must be adaptable enough to adapt to these changes while maintaining the coherence of the design. Another challenge is to balance personalization with universal design principles (such as accessibility, sustainability, and aesthetic consistency), avoid over-customization and ignore widely recognized design standards, and ensure that the final design meets functional, ethical, and social expectations.

In summary, personalized algorithms are changing the design process by generating spaces that are highly aligned with user preferences, lifestyles, and emotional needs. These algorithms optimize the functionality and emotional resonance of designs, ensuring that spaces reflect the user's identity while enhancing their overall experience. Despite the challenges of balancing privacy and design, their potential in creating meaningful, user-centric environments presents a huge opportunity for the future of design.

\section{Case Studies and Applications}

\subsection{Case Study 1: Office Space Design}
In one office remodeling project, machine learning models generated multiple layout options by analyzing a large amount of relevant data such as employee activities, environmental factors (such as lighting, air quality), and space utilization. These options prioritized practical needs such as workspace allocation, energy efficiency, and traffic flow, effectively optimizing space use and ensuring the proper distribution of desks, meeting rooms, and common areas throughout the office floor.

Although the initial layout design was highly functional, it lacked an important emotional component and failed to fully focus on employee comfort, mental health, and the creation of a positive work environment. This exposed a key limitation of machine-generated design: although machine learning models can effectively optimize quantifiable indicators such as space utilization, they tend to ignore subjective design factors that are critical to the human experience.
To this end, the designer worked with the client to adjust the layout to incorporate features that align with cultural values and lifestyle preferences. The improved design added common spaces such as lounges and dining areas, while maintaining a sense of privacy through partitions and hidden corners. The addition of warm colors, soft textures, and personalized decorations further created a warm atmosphere. The final design not only improved spatial efficiency, but also met emotional and cultural needs.

\begin{figure*}[t]
    \centering
    \includegraphics[width=\textwidth]{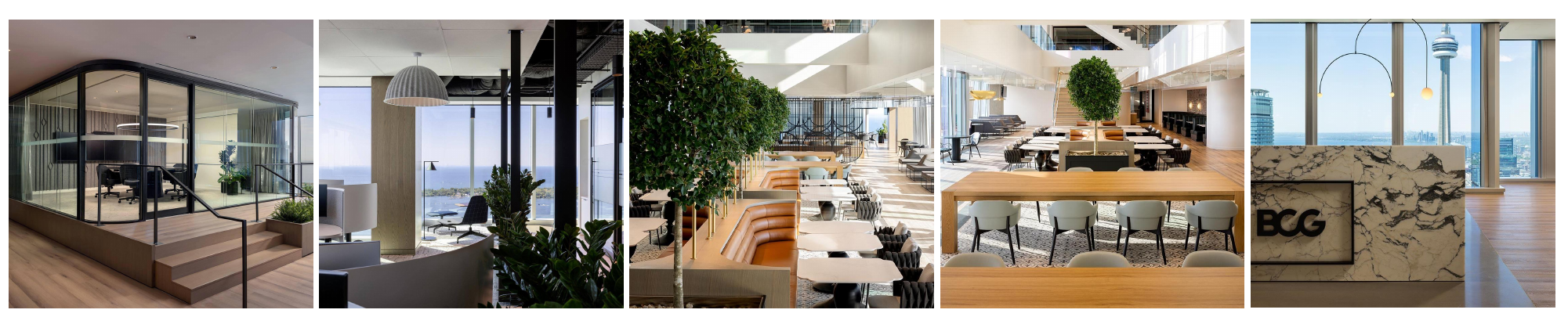}
    \caption{Case 1: the major office redesign project for Google's new office space in London.}
    \label{fig:case_8}
\end{figure*}

In the redesign of Google's new London office space (Figure \ref{fig:case_8}), machine learning models generated a series of efficient layout solutions by analyzing employee activity patterns, traffic flow, and environmental factors. These solutions optimize space use and energy efficiency, allowing Google to accommodate a large number of employees in a compact and efficient layout, but they also lack emotional and psychological considerations. To this end, Google worked with HOK, a well-known architectural firm known for its human-centered design, and combined employee feedback to introduce biophilic design elements (such as indoor plants, green walls, and natural wood) as well as ergonomic furniture, personalized workstations, and leisure areas. These improvements not only improve employee comfort, but also promote creativity and collaboration, making the office space more dynamic and health-oriented.

This case demonstrates the power of combining machine learning with human creativity: machine learning ensures functional efficiency, while the designer's professional insights give the space emotional resonance and personalized experience, ultimately achieving a balance between functionality and humanistic care.

\subsection{Case Study 2: Residential Space Design}
In a residential design project in a densely populated city, machine learning was used to optimize the spatial layout of a small apartment. The algorithm considered factors such as room size, user preferences, and environmental sustainability to generate solutions that maximized spatial efficiency. However, the final design needed further refinement to meet the cultural needs of the client. By incorporating human feedback, such as the desire for a warm atmosphere and respect for traditional family structures, the design increased public space and privacy elements, ultimately creating a practical and culturally appropriate home that met the family's physical and emotional needs.

\begin{figure*}[t]
    \centering
    \includegraphics[width=\textwidth]{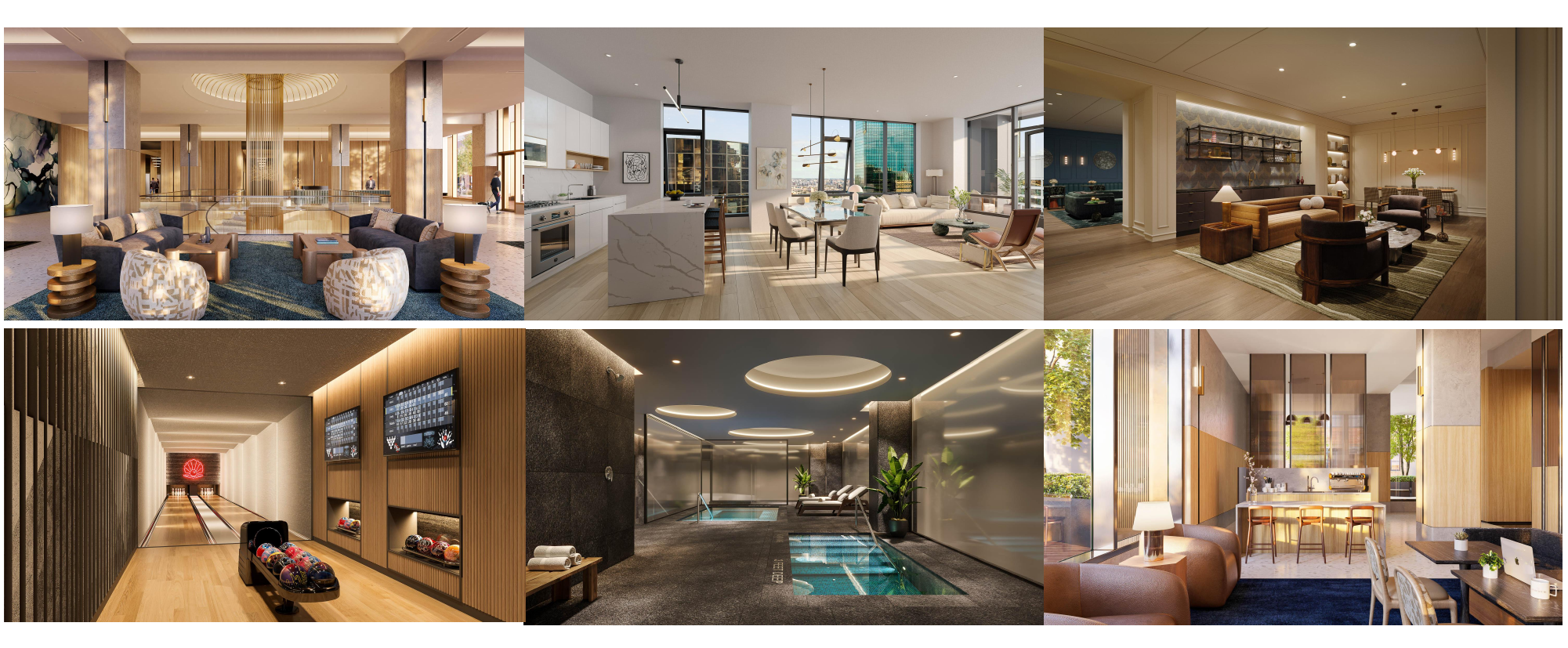}
    \caption{Case 2: the project for a small apartment in New York, SpaceIQ, a workplace management and space planning company, used machine learning to optimize the apartment layout.}
    \label{fig:case_9}
\end{figure*}

For a small apartment project in New York, workplace management and space planning company SpaceIQ used machine learning to optimize the layout (Figure \ref{fig:case_9}). The algorithm combined room size, user preferences, and sustainable building practices to generate multiple solutions for efficient use of space. Although the initial design performed well in terms of space utilization, it did not fully meet the emotional needs of the client. The family is from Japan and values privacy and family interaction in public spaces. To this end, SpaceIQ worked with Mosaic Design, a company specializing in culturally sensitive design, to adjust the layout to increase public areas, traditional tatami rooms, and secluded private spaces. Mosaic Design also created a welcoming atmosphere that matched the family’s expectations through soft textures, muted lighting, and warm earth tones. This design not only overcomes the spatial limitations of urban living, but also fully reflects the family’s emotional and cultural preferences.

\subsection{Case Study 3: Healthcare Facility Design}
In a medical facility design project, machine learning models generated efficient layout solutions based on patient flow, accessibility, and medical needs. However, these solutions lacked emotional warmth, which is critical in a medical environment centered on patient health. To this end, designers worked with medical professionals to incorporate elements such as natural light, soothing colors, and quiet areas to optimize the patient recovery environment. These adjustments have enabled the space to improve efficiency while promoting patient recovery, creating a balanced and compassionate medical environment.

\begin{figure*}[t]
    \centering
    \includegraphics[width=\textwidth]{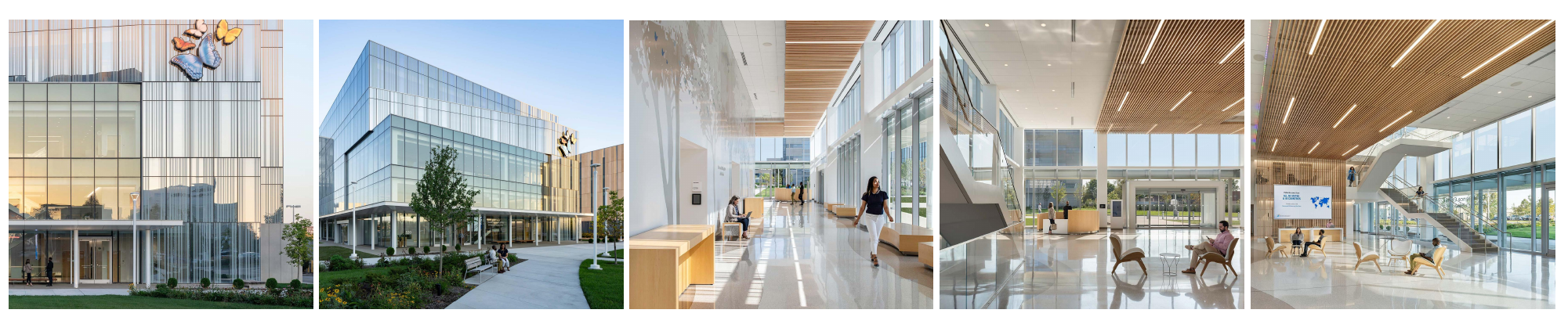}
    \caption{Case 3: the design of a new Cleveland Clinic patient room. Here, machine learning algorithms create highly optimized layouts for patient care areas based on patient flow data, accessibility needs, and medical workflow.}
    \label{fig:case_10}
\end{figure*}

Taking the design of the new wing of the Cleveland Clinic as an example, the machine learning algorithm generated a highly optimized care area layout (Figure \ref{fig:case_10}), ensuring that medical staff can quickly access patients, equipment, and medical records while minimizing congestion and improving overall operational efficiency. Although the layout is functionally efficient, it lacks a warm and healing atmosphere. The design team Gensler worked with medical professionals to introduce elements such as natural light, calming colors, and noise reduction strategies, and combined with patient feedback to add large windows, soundproofing measures, and quiet spaces. These improvements not only improve care efficiency, but also create a more compassionate recovery environment.

These examples show how machine learning combined with human expertise can create designs that are both functional and emotional. Machine learning optimizes spatial efficiency, energy use, and traffic flow, while human designers further refine the design by addressing emotional, cultural, and psychological needs. Whether in office spaces, residential environments, or healthcare facilities, this collaboration optimizes not only functionality but also creativity, well-being, and comfort.

By combining data-driven insights with human-centered design, companies like HOK, Mosaic Design, and Gensler are leading the design revolution, demonstrating how machine learning can drive innovation in the design process. In the future, design will leverage these technologies to create spaces that are both functional and personalized, adaptable, and emotionally resonant.

\section{Conclusion}

This paper proposes a framework for human-machine collaboration in spatial design, emphasizing the synergy between machine learning models and humanistic design principles. Although the role of machine learning in spatial design will continue to expand with technological advances, human creativity, intuition, and cultural awareness remain core elements for creating meaningful and impactful spaces. In the future, human-machine collaboration will continue to drive design innovation, ensuring that spaces meet both functional needs and evoke emotional resonance, thereby providing users with a more personalized and human experience.

\authorcontributions{For research articles with several authors, a short paragraph specifying their individual contributions must be provided. The following statements should be used ``Conceptualization, X.X. and Y.Y.; methodology, X.X.; software, X.X.; validation, X.X., Y.Y. and Z.Z.; formal analysis, X.X.; investigation, X.X.; resources, X.X.; data curation, X.X.; writing---original draft preparation, X.X.; writing---review and editing, X.X.; visualization, X.X.; supervision, X.X.; project administration, X.X.; funding acquisition, Y.Y. All authors have read and agreed to the published version of the manuscript.'', please turn to the  \href{http://img.mdpi.org/data/contributor-role-instruction.pdf}{CRediT taxonomy} for the term explanation. Authorship must be limited to those who have contributed substantially to the work~reported.}

\funding{Please add: ``This research received no external funding'' or ``This research was funded by NAME OF FUNDER grant number XXX.'' and  and ``The APC was funded by XXX''.}

\institutionalreview{Not applicable.}



\dataavailability{We encourage all authors of articles published in MDPI journals to share their research data. In this section, please provide details regarding where data supporting reported results can be found, including links to publicly archived datasets analyzed or generated during the study. Where no new data were created, or where data is unavailable due to privacy or ethical restrictions, a statement is still required. Suggested Data Availability Statements are available in section ``MDPI Research Data Policies'' at \url{https://www.mdpi.com/ethics}.}

\acknowledgments{In this section you can acknowledge any support given which is not covered by the author contribution or funding sections. This may include administrative and technical support, or donations in kind (e.g., materials used for experiments).}

\conflictsofinterest{Declare conflicts of interest or state ``The authors declare no conflicts of interest.'' Authors must identify and declare any personal circumstances or interest that may be perceived as inappropriately influencing the representation or interpretation of reported research results. Any role of the funders in the design of the study; in the collection, analyses or interpretation of data; in the writing of the manuscript; or in the decision to publish the results must be declared in this section. If there is no role, please state ``The funders had no role in the design of the study; in the collection, analyses, or interpretation of data; in the writing of the manuscript; or in the decision to publish the results''.}


\reftitle{References}


\bibliography{references}


%


\PublishersNote{}
\end{document}